\documentclass[12pt, letterpaper, notitlepage]{iopart}
\usepackage{graphicx,color}
\begin{document}
\newcommand{\kk}{{\bf k}}
\newcommand{\Q}{{\bf Q}}
\newcommand{\q}{{\bf q}}
\newcommand{\gk}{g_\textbf{k}}
\newcommand{\ee}{\tilde{\epsilon}^{(1)}_\textbf{k}}
\newcommand{\HH}{\mathcal{H}}

\title{Gap Symmetry and Stability Analysis in the Multi-Orbital Fe-Based Superconductors}
\author{Mark H. Fischer}
\address{%
Department of Physics, Cornell University, Ithaca, New York 14853, USA
}%
\ead{mark.fischer@cornell.edu}

\date{\today}

\begin{abstract}
The iron-based superconductors allow for a zoo of possible order parameters due to their orbital degrees of freedom. 
These order parameters are often written in an orbital basis for a microscopic analysis and a comprehensive symmetry classification. 
Unlike in standard single-band superconductors, where electrons of the same band - and hence, energy - are paired, a general order parameter in such a multi-orbital system also contains pairing of electrons belonging to different bands. 
As this corresponds to pairing of electrons of different energy, such order parameters are energetically less stable.
Here, we present a simple criterion for a stability analysis of the orbital part of the gap function based on the basic principle that electrons of equal energy are paired in the superconducting state.
This not only allows to find the most stable states, but also to identify the terms in a (tight-binding) Hamiltonian suppressing any given superconducting state.
Our approach thus allows to identify the minimal Hamiltonian necessary to compare competing instabilities.
\end{abstract}

\maketitle

%%%%%%%%%%%%%%%%%%%%%%%%%%%%%%%%%%%%%%%%%%%%%%%%
% Introduction
%%%%%%%%%%%%%%%%%%%%%%%%%%%%%%%%%%%%%%%%%%%%%%%%

\section{Introduction}
\label{sec:intro}
The symmetry classification of superconducting order parameters has proven to be a powerful tool in the context of unconventional superconductivity\cite{sigrist:1991}.
In single-band superconductors one distinguishes various gap functions $\Delta(\kk)$ according to their behavior under all the symmetry transformations of the generating point group of the crystal, i.e., the irreducible representation they belong to. 
While the interactions are the dominant factor for determining the pairing channel, many interactions allow for instabilities of different symmetry.
With only gap functions belonging to the same irreducible representation mixing, this allows to find each class' critical temperature and thus the dominant pairing channel. 
An interaction with dominant nearest-neighbor character on a square lattice, such as antiferromagnetic spin fluctuations, for example, allows for extended-$s$-, or $d$-wave (singlet) order parameters, which can be analyzed independently for the leading instability\cite{moriya:2000}.

In the Fe-based superconductors, several bands are involved in the low-energy physics. 
According to bandstructure calculations, the main contribution to the density of states at the Fermi level stems from Fe d orbitals hybridizing with As p orbitals\cite{singh:2008a, lebegue:2007}.
In most compounds, the bands crossing the Fermi energy have mainly $d_{xz}$ and $d_{yz}$ orbital character and thus, minimal tight-binding models consider only these two orbitals\cite{li:2008b, han:2008, raghu:2008}.
After back-folding of the Brillouin zone due to the out-of-plane positions of the As ions leading to a two-Fe unit cell, these models indeed yield an overall agreement with the experimentally found Fermi surfaces. However, at least the three $t_{2g}$ orbitals have to be considered in order to find an agreement in the Fermi velocities as well\cite{lee:2009a, graser:2009}.

This multi-orbital nature of the Fe-pnictides complicates any order parameter analysis:
The additional orbital degrees of freedom lead to an abundance of gap functions, which are now matrices in orbital space\footnote{This is similar to the spin-space matrices used when describing spin-triplet pairing\cite{mineev:1999}}. A symmetry classification of all possible gap functions with respect to the generating point group $D_{4h}$ of the Fe-As layer has been done for two-, as well as three-orbital models for example in Refs.~\cite{zhou:2008, wan:2009} and Ref.~\cite{daghofer:2010}, respectively. One difficulty arising with such an analysis is that the order parameter has to be written in orbital space, whereas the pairing of electrons in the BCS picture is best described in band space. Indeed, only pairing of electrons with the same energy, hence belonging to the same band, leads to a superconducting instability. While the analytical transformation into the band basis to analyze the stability of an order parameter is straight forward in a two-orbital description\cite{zhou:2008, wan:2009, moreo:2009}, the generalization to three or even more orbitals is not possible. Alternatively, the loss of condensation energy due to nodes can be used as a criterion for the stability of the various gap functions. This requires knowledge of all the nodes of a given gap function and also only allows for an even cruder first estimate.

In this work, we build on the aforementioned pairing of equal-energy electrons and present a simple criterion to analyze the stability of a gap with a given orbital structure. 
Simply speaking, we analyze whether a gap structure corresponds to a pure intra-band pairing, or whether electrons belonging to different bands are paired. For a systematic analysis, a comprehensive symmetry analysis of the free Hamiltonian and the order parameter, both matrices in orbital space, is necessary first. Our approach then also allows to identify the terms in the Hamiltonian that (partially) suppress the other gap structures. This has an important application in determining the minimal Hamiltonian needed to (numerically) compare different superconducting instabilities on equal footing. Note that such an analysis is unlike comparing extended-$s$- and $d$-wave paring in the single-band case, where always electrons belonging to the same band are paired. Here, the suppression is due to inter-band pairing, more comparable to the suppression of spin-singlet superconductivity due to a magnetic field. 

After introducing the general scheme in the next section, we apply it to the case of the iron-based superconductors. 
In order to construct the general tight-binding models, we start in Section \ref{sec:symmetry} with a few comments on the symmetry of the Fe-As layer and its generating point group $D_{4h}$. 
We then focus our discusion in Section~\ref{sec:2b} on a two-orbital description of the Fe-As layer and, with experiments strongly supporting spin-singlet pairing, we restrict our analysis on this pairing channel\cite{johnston:2010b, hirschfeld:2011}. The two-orbital model carries the advantage that we can gain additional analytical insight from the linearized gap equation.
Finally, we also comment on the three-orbital model in Section~\ref{sec:3b}. As a starting point for both models, we present a comprehensive symmetry analysis of the Hamiltonian and the gap structures including the two-Fe unit cell.
As an interesting result, we find that the extended-$s$ and $d_{x^2-y^2}$-wave pairing in the (two-Fe unit cell) three-orbital model is suppressed by the hybridization of the $d_{xz}$/$d_{yz}$ with the $d_{xy}$ orbitals, whereas this suppression is missing in general two-orbital models or when working with a one-Fe unit cell.

%%%%%%%%%%%%%%%%%%%%%%%%%%%%%%%%%%%%%%%%%%%%%%%%
% Symmetry & Model Construction
%%%%%%%%%%%%%%%%%%%%%%%%%%%%%%%%%%%%%%%%%%%%%%%%
\section{Symmetry and Stability in Multi-Orbital Systems}
\label{sec:sc}
\subsection{Single-Band Systems}
Before discussing the multi-orbital case, we give a brief overview over some symmetry aspects and notation for a superconductor in a single-band system. Only considering spin-singlet superconductivity, the mean-field Hamiltonian is a $2\times2$ matrix in Nambu space, $\vec{\Psi}_{\kk} = (c^{\phantom{\dag}}_{\kk\uparrow}, c^{\dag}_{-\kk\downarrow})$, given by
\begin{equation}
  \underline{\HH}^{\rm MF}(\kk) = \left(\begin{array}{cc}\HH^{\uparrow}_{0}(\kk) & \Delta(\kk) \\ \Delta^{*}(\kk) & -\HH^{\downarrow}_{0}(-\kk)\end{array}\right).
  \label{eq:nambu}
\end{equation}
$\HH^{\uparrow}_0(\kk)=\HH^{\downarrow}_{0}(\kk) = \xi_{\kk}$($=\xi_{-\kk}$) is the free hopping Hamiltonian\footnote{For simplicity, we only consider time-reversal and inversion symmetric systems.} and $\Delta(\kk) = \Delta \psi(\kk)$ is the superconducting order parameter or gap function. Since $\xi_{\kk}$ describes the non-interacting system, it has to transform trivially under time reversal and all the symmetry operations of the generating point group $\mathcal{G}$ of the crystal. From the transformation behavior of the single-particle operator under $g\in\mathcal{G}$, $g c_{\kk} = c_{R_g\kk}$, and $R_g$ the corresponding rotation matrix in momentum space,  we find $\xi_{R_g\kk} = \xi_{\kk}$ for all $g\in\mathcal{G}$.
In a similar way, the pair-wave functions $\psi(\kk)$ are symmetry classified in terms of the irreducible representations of $\mathcal{G}$, i.e., their behavior under $\psi(\kk)\mapsto\psi(R_g\kk)$.
Note that the Pauli principle requires the full gap function to change sign under the exchange of the two electrons, 
leading to the well-known requirement that an even wave function $\psi(\kk)$ is combined with spin-singlet pairing. 
Within BCS theory, the mechanism of superconductivity requires the pairing of electrons of equal energy. In the above Hamiltonian, the gap function always connects two electrons with the same energy, thus allowing for a superconducting instability for arbitrary small (attractive) interactions.

\subsection{Multi-Orbital Systems}
In a system with $n$ orbitals, the Hamiltonian and the (spin-singlet) gap function are $n\times n$ matrices in orbital space, leading to a mean-field Hamiltonian analogous to Eq.~\eref{eq:nambu} that is a $2n\times2n$ matrix,
\begin{equation}
  \underline{\HH}^{\rm MF}(\kk) = \left(\begin{array}{cc}\hat{\HH}_{0}(\kk) & \hat{\Delta}(\kk) \\ \hat{\Delta}^{\dag}(\kk) & -\hat{\HH}_{0}(\kk)\end{array}\right),
  \label{eq:nambun}
\end{equation}
where we have again assumed time-reversal and inversion symmetry and for simplicity omitted any spin-orbit coupling.
The Hamiltonian in orbital space is of the general form
\begin{equation}
  \hat{\HH}_{0}(\kk) = \sum_{a}\hat{\Lambda}^{a}h_{a}(\kk),
  \label{eq:hkk}
\end{equation}
where $\{\hat{\Lambda}^{a}\}$ is a basis for the $n\times n$ matrices. The Hamiltonian Eq.~\eref{eq:nambun} still has to transform trivially under all the symmetry operations of the generating point group $\mathcal{G}$. Since the symmetry transformations now act on the basis functions as $g \vec{c}_{\kk} = \hat{U}_{g}\vec{c}_{R_g\kk}$, the individual elements of the Hamiltonian \eref{eq:hkk} transform as 
\begin{equation}
  g[\hat{\Lambda}^a h_{a}(\kk)] = (\hat{U}_{g}^{\phantom{\dag}} \hat{\Lambda}^{a} \hat{U}^{\dag}_{g}) h_{a}(R_g\kk).
  \label{eq:transformH}
\end{equation}
For matrices and functions that transform as one-dimensional representations, simply the product of them has to transform trivially. For higher-dimensional representations, multiple matrices and momentum functions can be combined (Such an example is shown in Section \ref{sec:3b}). It is thus helpful to write the Hamiltonian as
\begin{equation}
  \hat{\HH}_{0}(\kk) = \HH_0^{id}(\kk) + \sum_{\alpha}\hat{\HH}^{\alpha}_{0}(\kk),
  \label{eq:hkka1g}
\end{equation}
where $\HH_{0}^{id}(\kk)$ is proportional to the $n\times n$ identity matrix and $\hat{\HH}^{\alpha}_{0}(\kk)$ are all the other smallest parts transforming trivially. 

The (spin-singlet) superconducting order parameter $\hat{\Delta}(\kk)$ can similarly be written as 
\begin{equation}
  \hat{\Delta}(\kk) = \sum_{a}\hat{\Lambda}^{a}\psi_{a}(\kk).
  \label{eq:gapfun}
\end{equation}
This order parameter can now describe the pairing of electrons of all possible orbitals. As for the free Hamiltonian, the full symmetry of the order parameter can be found from
\begin{equation}
  g[\hat{\Delta}(\kk)] = \sum_{a}(\hat{U}_{g}^{\phantom{\dag}} \hat{\Lambda}^{a} \hat{U}^{\dag}_{g})\psi_{a}(R_g\kk).
  \label{eq:transformD}
\end{equation}
The multi-orbital structure leads to a plethora of different gaps due to the additional degree of freedom. Most of these gaps will, however, lead to inter-band pairing and are thus less (or not at all) stable. After a brief discussion of the stability of general gap structures, we will see in Section \ref{sec:2b} for the two-orbital model and in Section \ref{sec:3b} for the three-band model an explicit symmetry classification and stability analysis of such order parameters.

\subsection{Stability of the Gap Structures}
In this section, we discuss a simple qualitative argument for the stability of gap structures and show that only the ones with $[\hat{\HH}_{0}(\kk),\hat{\Delta}(\kk)]=0$ are fully supported by the Hamiltonian.  
For a quantitative stability analysis of the terms in the Hamiltonian harming the superconducting state, one can for example resort to the linearized gap equation as done analytically for a two-band model in Section \ref{sec:lge}. 
The superconducting instability requires to pair electrons of equal and opposite momentum, and most importantly, equal energy.
For a gap structure that commutes with the orbital structure of the Hamiltonian, i.e. $[\hat{\HH}_{0}(\kk), \hat{\Delta}(\kk)]=0$, $\hat{\HH}_{0}(\kk)$ and $\hat{\Delta}(\kk)$ can both be simultaneously diagonalized by the same $n\times n$ matrix $\hat{U}_{\kk}$. The block-diagonal matrix with the two blocks $\hat{U}_{\kk}$ thus leads to a diagonal form of the mean-field equation~\eref{eq:nambun} with the gap only connecting states with equal energy, i.e., the gap function describes intra-band pairing only. 
If $[\hat{\HH}_{0}(\kk),\hat{\Delta}(\kk)]\neq 0$, we define the set $\mathcal{L} = \{\alpha| [\hat{\HH}^{\alpha}, \hat{\Delta}(\kk)]=0\}$ and write the Hamiltonian as 
\begin{equation}
  \hat{\HH}_{0}(\kk)=  \hat{\HH}_0^{id}(\kk) + \sum_{\alpha\in\mathcal{L}}\hat{\HH}_{0}^{\alpha}(\kk) + \sum_{\beta\notin\mathcal{L}}\hat{\HH}_{0}^{\beta}(\kk)
  \label{eq:}
\end{equation}
with  the first sum commuting with $\hat{\Delta}(\kk)$, while the second sum does not. 
If the Hamiltonian consisted of only the first two terms, the gap structure would describe intra-band pairing only and would be fully supported. The last sum, however, leads to inter-band contributions. The fact that there are terms in this last sum does, however, not necessarily lead to a full suppression of the instability.
As we will see explicitly for the two-band model, the stability instead depends on the respective energy scales of the two groups of terms. This scheme thus allows for a first estimate of the stability of all possible gap functions and also provides an insight into what terms in the non-interacting Hamiltonian harm a specific superconducting gap function. 

%%%%%%%%%
\section{Symmetry of the Fe-As Layer}
\label{sec:symmetry}
\begin{figure}[tb]
  \begin{center}
    \includegraphics{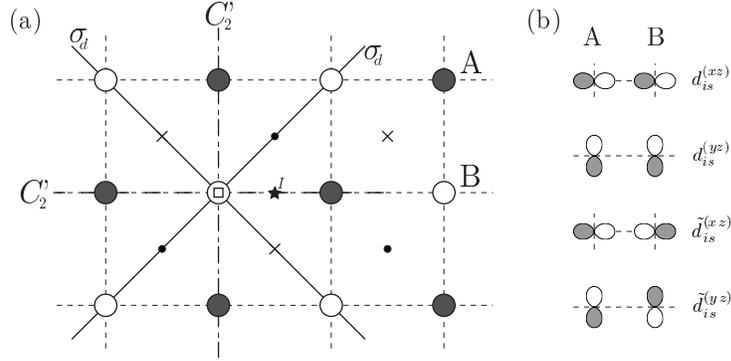}
  \end{center}
  \caption{(a) Cristal structure of the Fe-As layer with the two Fe sublattices $A$ (dark circles) and $B$ (white circles). The small crosses (dots) in the middle of the Fe squares are the As sites above (below) the plane. Also, the symmetry operations belonging to $D_{2d}$ are depicted with $C_2$ and $S_{4}$ being around the square on one of the Fe sites. Finally, the star between two iron sites denotes a center of inversion. (b) Orbital basis for the two-Fe unit cell.}
  \label{fig:unitcell}
\end{figure}

In order to apply the above scheme, we first recapitulate the symmetry properties of a single Fe-As plane\cite{lee:2009a, daghofer:2010}.
Figure \ref{fig:unitcell}(a) shows the structure of the Fe-As layer with its two-sublattice structure due to the out-of-plane positions of the As ions. The individual (two-Fe) unit cell is invariant under the following symmetry operations: The identity $E$, a two-fold rotation $C_{2}$ and two improper rotation $S_4$, both around the $z$ axis, two two-fold rotations $C_{2}'$ around the $x$ ($y$) axis, as well as two mirror planes $\sigma_{d}$ along the diagonals of the $xy$ plane, together building the point group $D_{2d} = \{E, C_2, 2C_2', 2S_4, 2\sigma_d\}$.
Note that the generating point group of the layer is isomorphic to $D_{4h}$, where all the elements $D_{4h}\setminus D_{2d}$ have to be combined with a lattice translation interchanging the two sublattices\footnote{A crystal with this property is called non-symmorphic.}. Taking this into account ensures having the inversion as a symmetry element of the layer, though not the unit cell. This is crucial for both, the construction of the Hamiltonian as well as the classification and analysis of the superconducting order parameter. 

\section{Two-Orbital Model}
\label{sec:2b}
\subsection{Hopping Hamiltonian}
\label{ssec:hop2}
In this section, we discuss the symmetry allowed terms of the multi-orbital hopping Hamiltonian based on the $d_{xz}$ and $d_{yz}$ orbitals with real-space operators $d^{(xz)}_{i s}$ and $d^{(yz)}_{is}$, respectively. We work with the two-Fe unit cell, and hence, additionally introduce the states $\tilde{d}^{(xz)}_{i s}$ and  $\tilde{d}^{(yz)}_{is}$, where the orbitals on the B sublattice are multiplied by $-1$ (see Fig.~\ref{fig:unitcell}(b)). In momentum space, this translates to having two operators $d^{(xz)}_{\kk s}$ / $d^{(xz)}_{\kk+\Q s}$ and $d^{(yz)}_{\kk s}$ / $d^{(yz)}_{\kk+\Q s}$, respectively with $\Q=(\pi,\pi)$. It is thus convenient to work in a tensor space of the $(d_{xz},d_{yz})$ orbital space and the $(\kk,\kk+\Q)$ momentum space.
The Hamiltonian can then be factorized in analogy to Eq.~\eref{eq:hkk} as
\begin{equation}
  \HH_{0} =\sum_{\kk, s}\vec{\mathcal{C}}_{\kk s}^{\,\dag} \Big[\sum_{a,b} \hat{\Sigma}^{ab}h_{ab}(\kk)\Big]\vec{\mathcal{C}}_{\kk s}^{\phantom{\dag}}
  \label{eq:hamgen}
\end{equation}
with $\vec{\mathcal{C}}_{\kk s} = (d^{(xz)}_{\kk s}, d^{(yz)}_{\kk s}, d^{(xz)}_{\kk+\Q s}, d^{(yz)}_{\kk+\Q s})^{T}$ and $\hat{\Sigma}^{ab} = \hat{\rho}^a\otimes\hat{\tau}^b$ is a tensor product of $\hat{\rho}^a$ and $\hat{\tau}^b$, the identity ($a,b=0$) and Pauli matrices ($a,b=1,2,3$) for the $(d_{xz}, d_{yz})$ and $(\kk, \kk+\Q)$ spaces, respectively. In the following, we separately analyze the behavior of $h_{ab}(\kk)$ and $\hat{\Sigma}^{ab}$ under all the symmetry transformations of the Fe-As layer. 
\begin{table}[bt]
  \centering
  \begin{tabular}{c|c|c|c|c|c|c|c|c}
    IR($D_{4h}$) & $A_{1g}$ & $A_{2g}$ & $B_{1g}$ & $B_{2g}$ & $A_{1u}$ & $A_{2u}$ & $B_{1u}$ & $B_{2u}$ \\[0.5ex]
    \hline
    $\hat{\Sigma}^{ab}$ & $\hat{\Sigma}^{00}, \hat{\Sigma}^{03}$ &$\hat{\Sigma}^{20}, \hat{\Sigma}^{23}$ &$\hat{\Sigma}^{30}, \hat{\Sigma}^{33}$ &$\hat{\Sigma}^{10}, \hat{\Sigma}^{13}$ &$\hat{\Sigma}^{31}, \hat{\Sigma}^{32}$& $\hat{\Sigma}^{11}, \hat{\Sigma}^{12}$ &$\hat{\Sigma}^{01}, \hat{\Sigma}^{02}$  & $\hat{\Sigma}^{21}, \hat{\Sigma}^{22}$
  \end{tabular}
  \caption{Classification of the orbital basis matrices $\hat{\Sigma}^{ab} = \hat{\rho}^{a}\otimes\hat{\tau}^{b}$ for a two-orbital $d_{xz}$/$d_{yz}$ basis. Note that the behavior under inversion ($u/g$) solely depends on the momentum part ($\hat{\tau}^{b}$), while the behavior under the symmetry transformations of the unit cell ($D_{2d}$) is given by the orbital part $\hat{\rho}^a$ (see \ref{app:transform2}).} 
  \label{tab:basism}
\end{table}

Table \ref{tab:basism} shows the classification of all the matrices $\hat{\Sigma}^{ab}$ with respect to $D_{4h}$ (see \ref{app:transform2} for details). As can be seen from this table,  the `momentum' part $\hat{\tau}^{b}$ determines the behavior under inversion ($u/g$). This has important consequences for the possible terms that can appear in the Hamiltonian: Momentum functions that are even (odd) under inversion can only be combined with $\hat{\tau}^{0}$ and $\hat{\tau}^{3}$ ($\hat{\tau}^1$ and $\hat{\tau}^2$). 
Table \ref{tab:basisf} summarizes all the hopping factors $h_{ab}(\kk)$ from up to third-nearest-neighbor hopping together with the $\hat{\tau}$ matrices in momentum space they have to be combined with. By construction, also the transformation $\kk\mapsto \kk+\Q$ has to be considered, such that for example the basis functions $(\cos k_x \pm \cos k_y)=-[\cos(k_x+\pi)\pm \cos (k_y+\pi)]$ have to be combined with $\hat{\tau}^3$.
\begin{table}[tb]
    \centering
\begin{tabular}{l c c}
  \hline\hline
  IR& $\hat{\tau}^{0}$ & $\hat{\tau}^{3}$\\
  \hline
  $A_{1g}$ \;\;& $1$, $\cos k_x\cos k_y$, $\cos 2 k_x + \cos 2 k_y$ & $\cos k_x + \cos k_y$\\[0.5ex]
  $B_{1g}$ & $\cos 2 k_x - \cos 2 k_y$ & $\cos k_x - \cos k_y$\\[0.5ex]
  $B_{2g}$& $\sin k_x \sin k_y$  & - \\[1.5ex]
  \hline
  &  $\hat{\tau}^{1}$ & $\hat{\tau}^{2}$\\
  \hline
  $E_{u}$ & $\{\sin k_x\cos k_y, \sin k_y \cos k_x\}$ & $\{\sin k_x, \sin k_y\}$\\[0.5ex]
  & $\{\sin 2k_x, \sin 2k_y\}$\\[0.5ex]
  \hline\hline
\end{tabular}
\caption{The basis functions for hopping terms $h_{ab}(\kk)$ up to third-nearest neighbors classified in terms of irreducible representations of $D_{4h}$. The functions are divided into the ones that change sign for $\kk\mapsto\kk+\Q$, thus allowing for combinations with $\hat{\tau}^{3}$ ($\hat{\tau}^2$) in the Hamiltonian, and those that do not, thus allowing for combinations with $\hat{\tau}^{0}$ ($\hat{\tau}^{1}$).}
    \label{tab:basisf}
\end{table}

We are now in the position to construct the Hamiltonian by requiring it to transform as the trivial representation $A_{1g}$. Since only one-dimensional representations appear in Table \ref{tab:basism}, two elements of the same irreducible representation have to be combined, e.g. $\cos k_x\cos k_y$ with $\hat{\Sigma}^{00}$. 
The Hamiltonian reads
\begin{equation}
  \HH_{0} = \sum_{\kk, s}\vec{\mathcal{C}}_{\kk s}^{\,\dag}[\hat{\Sigma}^{00}h_{00}(\kk) +\hat{\Sigma}^{03}h_{03}(\kk)  +  \hat{\Sigma}^{30} h_{30}(\kk) +  \hat{\Sigma}^{33}h_{33}(\kk) +  \hat{\Sigma}^{10}h_{10}(\kk)]\vec{\mathcal{C}}_{\kk s}^{\phantom{\dag}},
  \label{eq:ham0}
\end{equation}
with $h_{00}(\kk) = 4 t_{2s} \cos k_x \cos k_y + 2 t_{3s} (\cos 2k_{x} + \cos 2k_{y})-\mu$, $h_{03}(\kk) = 2t_{1s}(\cos k_x + \cos k_y)$, $h_{30}(\kk)=2t_{3d}(\cos 2k_x - \cos 2k_y)$, $h_{33}(\kk) = 2t_{1d} (\cos k_x - \cos k_y)$, and $h_{10}(\kk) = 4 t_{2d} \sin k_x \sin k_y$. Note that these hopping terms correspond to isotropic hopping along the $x$ and $y$ axis ($h_{00}(\kk)$ and $h_{03}(\kk)$), anisotropic hopping with respect to the axes ($h_{30}(\kk)$ and $h_{33}(\kk)$), as well as inter-orbital hopping ($h_{10}(\kk)$).
The Hamiltonian \eref{eq:ham0} has 4 bands with dispersions
\begin{equation}
  \xi^{(a)}_{\pm}(\kk) = \epsilon_{+}^{(a)}(\kk) \pm \sqrt{[\epsilon_{-}^{(a)}(\kk)]^2 + [h_{10}(\kk)]^2},
  \label{eq:xi0}
\end{equation}
with $\epsilon_{+}^{(1/2)}(\kk) = h_{00}(\kk) \pm h_{03}(\kk)$, $\epsilon_{-}^{(1/2)}(\kk) = h_{30}(\kk) \pm h_{33}(\kk)$, and $a = (1,2)$ for the subspace of $(\kk, \kk+\Q)$.
Note that no terms transforming trivially can be constructed containing the matrices $\hat{\tau}^1$ or $\hat{\tau}^{2}$ and thus, 
there is no allowed coupling terms between $\kk$ and $\kk + \Q$. The first Brillouin zone is thus only artificially folded down to the two-Fe zone. Terms that explicitly fold the Brillouin zone in such a two-orbital description necessarily involve spin-orbit coupling.  When sketching the gap functions below, we thus consider only one of the two subspaces for simplicity. 
 
\begin{figure}[tb]
  \centering
    \includegraphics{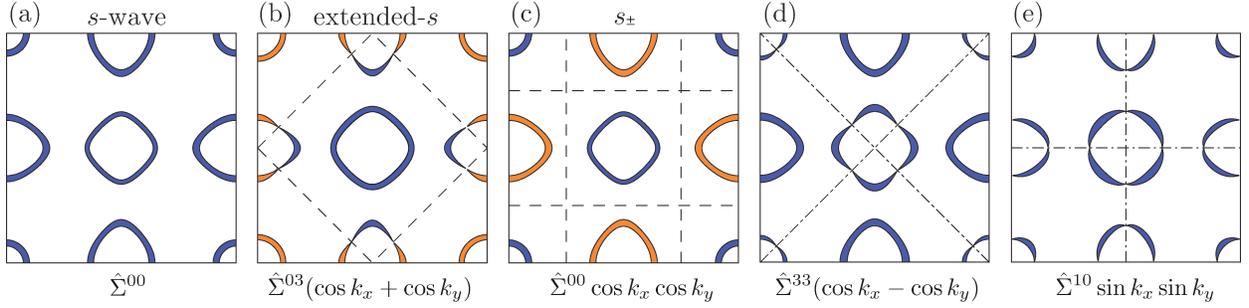}
    \caption{Sketch of the possible spin-singlet gap structures in momentum space with the full $A_{1g}$ symmetry. (a) shows the standard $s$-, (b) extended-$s$- and (c) $s_{\pm}$-wave gap functions. (d) and (e) are possible fully symmetric gap functions with band- and momentum structure belonging to $B_{1g}$ and $B_{2g}$, respectively. The two colors depict the sign change of the gap and the dashed (dotted) lines the nodes of the momentum (orbital) part of the gap. Note that for the gap functions belonging to $A_{1g}$ these two lines have to fall together.}
  \label{fig:A1gaps}
\end{figure}

\subsection{Gap Classification and Stability}
Only considering spin-singlet gap functions already allows for a large variety of different gap functions (see e.g. \cite{zhou:2008, wan:2009}). The combination of all the matrices from Table~\ref{tab:basism} with momentum functions from Table~\ref{tab:basisf} allows for 54 different order parameters and we thus discuss only some examples. Figure \ref{fig:A1gaps} sketches the $A_{1g}$ gap functions with momentum parts going up to next-nearest neighbors and only showing one of the momentum subspaces in the one-Fe Brillouin zone. In addition to the conventional $s$-wave, extended-$s$ and $s_{\pm}$ gap functions, given by
\begin{eqnarray}
  \hat{\Delta}(\kk) &=& \hat{\Sigma}^{00},\label{eq:A1a1}\\
  \hat{\Delta}(\kk) &=& \hat{\Sigma}^{03}(\cos k_x + \cos k_y),\label{eq:A1a2}\\
  \hat{\Delta}(\kk) &=& \hat{\Sigma}^{00} \cos k_x \cos k_y,
  \label{eq:A1a3}
\end{eqnarray}
also gap functions with gap nodes on the Fermi surface around the $\Gamma$ point, namely 
\begin{eqnarray}
  \hat{\Delta}(\kk) &=& \hat{\Sigma}^{33}(\cos k_x - \cos k_y),\label{eq:a1b}\\ 
  \hat{\Delta}(\kk) &=& \hat{\Sigma}^{10}\sin k_x \sin k_y
  \label{eq:a1c}
\end{eqnarray}
are shown. 
\begin{figure}[tb]
  \centering
    \includegraphics{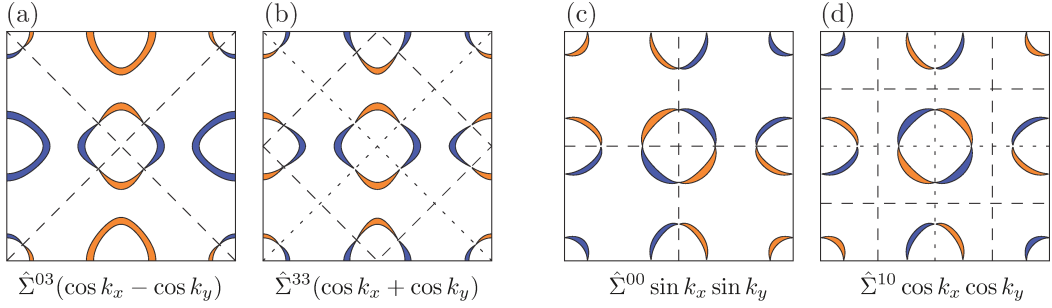}
    \caption{Momentum space representations of the gap functions with $B_{1g}$ [(a), (b), and (c)] and $B_{2g}$ [(d) and (e)] symmetry. (a) is the standard $d_{x^2-y^2}$- and (d) the $d_{xy}$ wave gap. (b), (c), and (e) have an $A_{1g}$ momentum dependence combined with a $B_{1g}$ and $B_{2g}$ band dependence, respectively. In addition to the nodes dictated by their momentum part (dashed lines), they thus also have nodes coming from their orbital structure (dotted lines).}
  \label{fig:Bgaps}
\end{figure}

Figure~\ref{fig:Bgaps} shows gap functions that transform as $B_{1g}$ and $B_{2g}$. These are, first, the so-called $d_{x^2-y^2}$-wave gap (Fig.~\ref{fig:Bgaps}(a))
\begin{equation}
  \hat{\Delta}(\kk) = \hat{\Sigma}^{03}(\cos k_x - \cos k_y)
  \label{eq:dx2-y2}
\end{equation}
and $d_{xy}$-wave gap (Fig.~\ref{fig:Bgaps}(d))
\begin{equation}
  \hat{\Delta}(\kk) = \hat{\Sigma}^{00}\sin k_x \sin k_y.
  \label{eq:dxy}
\end{equation}
In addition, three gap structures are shown with a `trivial' momentum part in combination with a $B_{1g}$ (Fig.~\ref{fig:Bgaps}(b), (c)) and $B_{2g}$ (Fig.~\ref{fig:Bgaps}(e)) orbital part, respectively,
\begin{eqnarray}
  \hat{\Delta}(\kk) &=& \hat{\Sigma}^{33}(\cos k_x + \cos k_y),\label{eq:b1b}\\ 
  \hat{\Delta}(\kk) &=& \hat{\Sigma}^{30}\cos k_x \cos k_y,\label{eq:bdpm}\\
  \hat{\Delta}(\kk) &=& \hat{\Sigma}^{10}\cos k_x \cos k_y.
  \label{eq:b2b}
\end{eqnarray}
The gap structures with non-trivial orbital part in Eqs.~\eref{eq:a1b} and \eref{eq:a1c}, as well as \eref{eq:b1b} - \eref{eq:b2b} have not only gaps due to their momentum dependence (dashed lines), but also their orbital dependence (dotted lines). Note, however, that here the former ones are not protected by symmetry and thus any admixture of other gap structures belonging to the same irreducible representation can shift or open these nodes. Nevertheless, these gaps by themselves are not fully gapped and hence, in general less stable.

The fact that the gaps with non-trivial orbital structure have (additional) nodes and thus a reduced superconducting condensation energy already points toward less stable superconductivity. Even without knowledge of these gap nodes, we can analyze the stability of the above gap structures using the criterion introduced in Section~\ref{sec:sc}. All the gap structures that transform trivially in the two subspaces $(\kk, \kk+\Q)$, namely the ones defined in Eqs.~\eref{eq:A1a1}-\eref{eq:A1a3}, \eref{eq:dx2-y2}, and \eref{eq:dxy}, commute with the free Hamiltonian \eref{eq:ham0}. These gap structures are thus intra-band only and fully supported. However, the gap structures with non-trivial orbital parts do not commute with the Hamiltonian \eref{eq:ham0}. The gap structures defined in Eqs.~\eref{eq:a1b}, \eref{eq:b1b}, and \eref{eq:bdpm} do not commute with the term $\hat{\Sigma}^{10} h_{10}(\kk)$ in the Hamiltonian \eref{eq:ham0}, while they commute with all the other terms. This means that these gap functions are suppressed by inter-orbital hopping. The gap structures in Eqs.~\eref{eq:a1c} and \eref{eq:b2b}, on the other hand, do not commute with the terms $\hat{\Sigma}^{30}h_{30}(\kk)$ and $\hat{\Sigma}^{33}h_{33}(\kk)$, i.e., they are suppressed by the `one-dimensional nature' of the $d_{xz}$ / $d_{yz}$ bands.  

Finally, we comment on the peculiar gap structures that combine an odd momentum function with spin-singlet pairing. In order to obey the Pauli principle, either $\hat{\rho}^2$ or $\hat{\tau}^2$ has to appear in the decomposition of $\hat{\Sigma}^{ab}$. Examples are
\begin{equation}
  \hat{\Delta}(\kk) = \hat{\Sigma}^{02} \sin k_x,
  \label{eq:oddsinglet1}
\end{equation}
belonging to the two-dimensional representation $E_g$, or
\begin{equation}
  \hat{\Delta}(\kk) = \hat{\Sigma}^{23} \sin k_x
  \label{eq:oddsinglet2}
\end{equation}
belonging to $E_u$. The former gap structure corresponds to finite-momentum pairing in the one-Fe zone and it is thus suppressed. The latter gap structure only commutes with the terms proportional to $\hat{\Sigma}^{00}$ and $\hat{\Sigma}^{03}$ in the Hamiltonian \eref{eq:ham0}. From our discussion so far, we hence also expect this gap structure to be suppressed. Calculating the linearized self-consistency equation in the next section, we can explicitly analyze how the various terms influence the pairing instability for the above gap structures.

\subsection{Linearized Gap Equation}
\label{sec:lge}
In this section, we evaluate the linearized self-consistency equation for the gap structures to see explicitly how the criterion introduced in Section \ref{sec:sc} works out.
For this purpose, we introduce a spin-singlet pairing interaction of the general form
\begin{equation}
  \HH' = \frac{1}N \sum_{\kk, \kk'} V_{\alpha\beta,\mu\nu}(\kk, \kk')\mathcal{C}_{\alpha\kk \uparrow}^{\dag}\mathcal{C}_{\beta-\kk \downarrow}^{\dag}\mathcal{C}_{\mu -\kk' \downarrow}^{\phantom{\dag}}\mathcal{C}_{\nu\kk' \uparrow}^{\phantom{\dag}}
  \label{eq:int0}
\end{equation}
(We use the convention of summing over repeated indices).
In the framework of the Gor'kov equations, we find the linearized gap equation
\begin{equation}
  \Delta_{\alpha\beta}(\kk) = -T\sum_{\omega_n}\sum_{\kk'}V_{\alpha\beta,\mu\nu}(\kk, \kk') \times\Big[\hat{G}_{0}(\kk', \omega_n)\hat{\Delta}(\kk')\hat{G}_{0}^{T}(-\kk', -\omega_n)\Big]_{\mu\nu},
  \label{eq:linself}
\end{equation}
where $\hat{G}_0(\kk, \omega_n) = [i\omega_n \hat{\Sigma}^{00} - \hat{\HH}_0(\kk)]^{-1}$ is the non-interacting Green's function with $\omega_n = (2n+1)\pi T$ the fermionic Matsubara frequencies. This equation determines the critical temperature $T_c$ for a given gap structure, where the largest $T_c$ belongs to the leading instability.
Parametrizing the interaction similar to Eq.~\eref{eq:gapfun},
\begin{equation}
  V_{\alpha\beta,\mu\nu}(\kk, \kk')=\sum_{a,b}\sum_{l}v^{(l)}_{ab}\Big[\hat{\Sigma}^{ab}\psi^{(l)}_{ab}(\kk)\Big]_{\alpha\beta} \times\Big[\hat{\Sigma}^{ab}\psi^{(l)}_{ab}(\kk)\Big]_{\mu\nu},
  \label{eq:int-param}
\end{equation}
with $\psi_{ab}^{(l)}(\kk)$ the individual basis functions from Table \ref{tab:basisf}, we find the self-consistency equation for each individual gap structure. 

We neglect here all possible gap mixing and only focus on the diagonal elements. We first consider gap structures that commute with the Hamiltonian, namely the ones with a `trivial' structure $\hat{\Sigma}^{00}$ and $\hat{\Sigma}^{03}$, respectively, given in Eqs.~\eref{eq:A1a1}-\eref{eq:A1a3}, \eref{eq:dx2-y2}, and \eref{eq:dxy}.
After summation over the Matsubara frequencies, the linearized gap equation for each gap reads 
\begin{equation}
  1 = -v^{(l)}_{ab}\sum_{\kk'}\sum_{a}\sum_{\alpha=\pm}\frac{[\psi^{(l)}_{ab}(\kk')]^2}{2 \xi^{a}_{\alpha}(\kk')}\tanh\Big(\frac{\xi^{a}_{\alpha}(\kk')}{2T}\Big),
  \label{eq:selfconA1g}
\end{equation}
where $ab = 00, 03$ and $\psi^{(l)}_{ab}(\kk)$ is the momentum part of the respective gap. In this equation, every state with energy close to the Fermi energy adds to the instability due to the intra-band (equal-energy) character of the pairing with an instability even for an infinitesimal attractive interaction due to the divergence for $T\rightarrow 0$. As expected, these gap structures are thus fully supported by the hopping Hamiltonian.  

The non-trivial gap structures from Eqs.~\eref{eq:a1b} and \eref{eq:b1b} (and analogously for Eq.~\eref{eq:bdpm}) yield a self-consistency equation of the form
\begin{eqnarray}
  1 &=& -v^{(l)}_{33}\sum_{\kk',a, \alpha}\Big\{\frac{[\varepsilon_{-}^a(\kk')]^2}{[h_{10}(\kk')]^2 + [\varepsilon_{-}^a(\kk')]^2}\frac{[\psi^{(l)}_{33}(\kk')]^2}{2 \xi^{a}_{\alpha}(\kk')}\tanh\Big(\frac{\xi^{a}_{\alpha}(\kk')}{2T}\Big)\nonumber\\
  &&+ \frac{[h_{10}(\kk')]^2}{[h_{10}(\kk')]^2 + [\varepsilon_{-}^{(a)}(\kk')]^2}\frac{[\psi^{(l)}_{33}(\kk')]^2}{2\varepsilon_{+}^{a}(\kk')}\tanh\Big(\frac{\xi^{a}_{\alpha}(\kk')}{2T}\Big)\Big\}.
  \label{eq:selfcon33}
\end{eqnarray}
Note that the first term on the right-hand side again corresponds to intra-band pairing with the $T\rightarrow 0$ divergence, whereas the second term comes from inter-band pairing and has no such divergence. Hence, this self-consistency equation still allows for a solution for an infinitesimal interaction, however, $T_c$ is suppressed. As expected from the discussion in the previous section, the ratio $h_{10}(\kk) / \varepsilon_{-}^{a}(\kk)$ determines the strength of the instability. 
Analogously, we find
\begin{eqnarray}
  1 &=& -v^{(l)}_{10}\sum_{\kk',a, \alpha}\Big\{\frac{[h_{10}(\kk')]^{2}}{[h_{10}(\kk')]^2 + [\varepsilon_{-}^{a}(\kk')]^2}\frac{[\psi^{(l)}_{10}(\kk')]^2}{2 \xi^{a}_{\alpha}(\kk')}\tanh\Big(\frac{\xi^{a}_{\alpha}(\kk')}{2T}\Big)\nonumber\\
  &&+\frac{[\varepsilon_{-}(\kk')]^2}{[h_{10}(\kk')]^2 + [\varepsilon_{-}^{a}(\kk')]^2}\frac{[\psi^{(l)}_{10}(\kk')]^2}{2\varepsilon_{+}^{a}(\kk')}\tanh\Big(\frac{\xi^{a}_{\alpha}(\kk')}{2T}\Big)\Big\}
  \label{eq:selfcon10}
\end{eqnarray}
for the gap functions from Eqs.~\eref{eq:a1c} and \eref{eq:b2b} with the same structure as Eq.~\eref{eq:selfcon33}.
Finally, for the gap structure given in Eq.~\eref{eq:oddsinglet2} we find
\begin{equation}
  1 = -v^{(l)}_{23}\sum_{\kk'}\sum_{a}\sum_{\alpha=\pm}\frac{[\sin k'_x]^2}{2 \varepsilon^{a}_{+}(\kk')}\tanh\Big(\frac{\xi^{a}_{\alpha}(\kk')}{2T}\Big).
  \label{eq:selfconEg}
\end{equation}
This gap function is only describing inter-band pairing and thus has no $T\rightarrow0$ divergence and associated instability for an arbitrarily small interaction any more.

%%%%%%%%%%%%%%%%%%%%%%%%%%%%%%%%%%%%%%%%%%%%%%%%
% 3 Bands
%%%%%%%%%%%%%%%%%%%%%%%%%%%%%%%%%%%%%%%%%%%%%%%%
\section{Three-Orbital Model}
\label{sec:3b}
\begin{table}[bt]
    \centering
\begin{tabular}{c|c}
  IR($D_{2d}$) & $\hat{\lambda}^i$\\
  \hline
  \;\;$A_{1}$ \;\; & $\hat{\lambda}^0$, $\hat{\lambda}^{8}$\\
  \;\;$A_{2}$ \;\; & $\hat{\lambda}^2$\\
  \;\;$B_{1}$ \;\; & $\hat{\lambda}^3$\\
  \;\;$B_{2}$ \;\; & $\hat{\lambda}^1$\\
  \;\;$E$ \;\; & $\{\hat{\lambda}^4, \hat{\lambda}^6\}$, $\{\hat{\lambda}^{5}, \hat{\lambda}^7\}$
\end{tabular}
\caption{Classification of the Gell-Mann matrices in terms of irreducible representations of $D_{2d}$. The momentum part $\hat{\tau}^b$ then determines how the matrices $\hat{\Lambda}^{ab} = \hat{\lambda}_{a}\otimes\hat{\tau}_{b}$ behave under inversion and thus what irreducible representation they belong to (see \ref{app:transform2} for details).}
    \label{tab:basism3}
\end{table}
We now turn to the three-orbital model, where in addition to the $d_{xz}$ and $d_{yz}$ orbitals, also the $d_{xy}$ orbital is introduced. This is necessary for an accurate description of the low-energy electronic structure and we hence briefly repeat the analysis of the symmetry and stability of the gap functions.  
The Hamiltonian can again be factorized as
\begin{equation}
  \HH_{0} =\sum_{\kk, s}\vec{\mathcal{C}}_{\kk s}^{\,\dag} [\sum_{a,b} \hat{\Lambda}^{ab}h_{ab}(\kk)]\vec{\mathcal{C}}_{\kk s}^{\phantom{\dag}},
  \label{eq:hamgen3}
\end{equation}
where now, 
\begin{equation}
  \vec{\mathcal{C}}_{\kk s} = (d^{(xz)}_{\kk s}, d^{(yz)}_{\kk s}, d^{(xy)}_{\kk s},  
  d^{(xz)}_{\kk+\Q s}, d^{(yz)}_{\kk+\Q s}, d^{(xy)}_{\kk+\Q s})^T.
  \label{eq:3vec}
\end{equation}
We have introduced the matrices $\hat{\Lambda}^{ab} = \hat{\lambda}^a\otimes\hat{\tau}^b$, with $\hat{\tau}^{b}$ as before the matrices in $\{\kk, \kk+\Q\}$ space and the orbital basis is given in terms of the ($3\times3$) Gell-Mann matrices $\hat{\lambda}^{a}$ (see \ref{app:gell-mann}).
In analogy to Section \ref{ssec:hop2}, we analyze the behavior of these matrices under the symmetry operations of the generating point group. For simplicity, Table \ref{tab:basism3} only lists the classification of the Gell-Mann matrices. The full matrix $\hat{\Lambda}^{ab}=\hat{\lambda}^a\otimes\hat{\tau}^b$ then transforms like $\hat{\lambda}^a$ under $D_{2d}$ and is even (odd) under inversion for $b=0,3$ ($b=1,2$), c.f., Table \ref{tab:basism}.
All the matrix elements only connecting the $d_{xz}$ and $d_{yz}$ orbitals are (with the appropriate matrices) the same as for the two-orbital case. For the diagonal part of $d_{xy}$, we simply add the terms
\begin{eqnarray}
 &&\frac13(\hat{\Lambda}^{03} - \sqrt{3}\hat{\Lambda}^{83}) 2t_{xy} (\cos k_x + \cos k_y)\\
 &&+\frac{1}3(\hat{\Lambda}^{00} - \sqrt{3}\hat{\Lambda}^{80}) (4t_{xy}'\cos k_x \cos k_y - \mu_{xy})
 \label{eq:addterm}
\end{eqnarray}
including nearest- and next-nearest-neighbor hopping and the chemical potential $\mu_{xy}$.
Due to the out-of-plane position of the As atoms, there is also an allowed hopping between the $d_{xz}$ / $d_{yz}$ and the $d_{xy}$ orbitals. This is a nearest-neighbor hopping of the form
\begin{equation}
  \hat{\Lambda}^{42}\sin k_x - \hat{\Lambda}^{62} \sin k_y
  \label{eq:connectionn2}
\end{equation}
and a next-nearest-neighbor hopping 
\begin{equation}
  \hat{\Lambda}^{51}\sin k_x\cos k_y - \hat{\Lambda}^{71} \sin k_y\cos k_x,
  \label{eq:connectionn3}
\end{equation}
both odd under inversion.
Note that these terms now are made from matrices / momentum functions that transform like the two-dimensional representation $E_{u}$ and can thus not be split further. To choose which $E_{u}$ matrices from Table \ref{tab:basism3} have to be used, time-reversal symmetry has to be considered, which requires $H_{0}^{ab}(\kk) = [H^{ab}_{0}(-\kk)]^*$. Further note that the Hamiltonian again separates into two blocks, with $(d^{(xz)}_{\kk s}, d^{(yz)}_{\kk s}, d^{(xy)}_{\kk+\Q s})$ and vice versa\footnote{This separation into two blocks can be understood from the behavior under a glide-plane transformation as noted by Ref.~\cite{lee:2009a}}.

With this hopping Hamiltonian, we can again analyze the stability of possible gap structures. Compared with the two-orbital case, there is one key difference: Even without spin-orbit coupling the $\kk$ and $\kk+\Q$ subspaces are connected through the hopping terms in Eqs.~\eref{eq:connectionn2} and \eref{eq:connectionn3}. As a result, the only spin-singlet gap function that has no inter-band contributions and hence is fully supported by the Hamiltonian has the trivial orbital structure $\hat{\Lambda}^{00}$. This has two important consequences: First, the most stable gap structure  has the same momentum dependence on all Fermi surfaces, i.e., there is a single gap function determining the gap everywhere in the Brillouin zone\cite{daghofer:2010}. Second, the orbital structure $\hat{\Lambda}^{00}$ only allows for functions $\psi(\kk)$ that transform trivially under $\kk \mapsto \kk+\Q$. This means that the gap function with extended-$s$ and d$_{x^2-y^2}$ symmetry with $\psi(\kk) = \cos k_x \pm \cos k_y$ are suppressed by the nearest-neighbor hopping between the d$_{xz}$ / d$_{yz}$ and the d$_{xy}$ orbitals.\footnote{Note that this suppression is present even without hybridization of the electron pockets discussed in Ref.~\cite{khodas:2012}.}
%%%%%%%%%%%%%%%%%%%%%%%%%%%%%%%%%%%%%%%%%%%%%%%%
% Discussion & Conclusion
%%%%%%%%%%%%%%%%%%%%%%%%%%%%%%%%%%%%%%%%%%%%%%%%
\section{Discussion and Conclusion}
In this work, we have introduced a simple criterion to analyze the stability of the orbital structure of superconducting order parameters in the context of multi-orbital systems: A vanishing commutator of the gap function with the free Hamiltonian. 
This argument is based on intra-band pairing, i.e, the equal-energy pairing of electrons in a weak-coupling approach. Depending on the structure of the interaction, the instability can therefore still occur in an orbital channel that is not fully supported by the Hamiltonian, especially with almost degenerate bands and within strong coupling. The analysis is thus only a first step and resembles the argument regarding the nodes in the superconducting gap due to non-trivial orbital structures and the associated loss of condensation energy. However, the introduced scheme additionally allows for the identification of the superconductivity-suppressing hopping terms. In the case of competing gap functions, this can help finding a minimal model for an unbiased description.

We have applied our scheme to the case of spin-singlet pairing in the Fe-based superconductors. Looking at all symmetry-allowed hopping terms in a three-orbital model for the two-Fe unit cell, we have found that only the trivial orbital structure proportional to the identity is fully supported by the system, i.e., only describes intra-band pairing. In particular, the gap functions with $\psi(\kk+\Q) = -\psi(\kk)$ are suppressed by the hopping terms connecting $\kk$ and $\kk+\Q$. 
Moreover, the gap functions belonging to $B_{1g}$ in a two-dimensional model have additional gap nodes due to their orbital structure along the folded Brillouin zone, as this symmetry belongs to $B_{2g}$ in this reduced Brillouin zone\cite{mazin:2011}. The gap function shown in Fig.~\ref{fig:Bgaps}(c) then corresponds to a `$d_{\pm}$' gap\cite{thomale:2009}, to give an example.
As the suppression and the additional nodes are missed considering two-orbital or general one-Fe-unit-cell models, our analysis emphasizes the importance of using at least a (two-Fe) three-orbital description when studying possible competing order parameters. Especially in the description of Fe-based superconductors with only electron pockets this could be relevant\cite{wang:2011b, das:2011, maier:2011}.

Finally, we comment on the generalization of the introduced scheme to spin-triplet superconductivity and the introduction of spin-orbit coupling. For simplicity, we have focused in this work on spin-singlet pairing and no spin-orbit coupling. For a generalization, the Nambu spinors have to be written in a way as to preserve the form of the mean-field Hamiltonian~\eref{eq:nambu}. The application of the commutation criterion is then again straight forward. In the case of a single-band system with time-reversal symmetry, for example, the appropriate basis is $\vec{\Psi}(\kk) = (c_{\kk\uparrow}, c_{\kk\downarrow}, c_{-\kk\downarrow}^{\dag}, -c_{-\kk\uparrow}^{\dag})^{T}$. The lifting of the degeneracy of the spin-triplet gap functions in non-centrosymmetric superconductors\cite{mineev:2012} is for example easily obtained in this way.

%%%%%%%%%%%%%%%%%%%%%%%%%%%%%%%%%%%%%%%%%%%%%%%%
% Appendix
%%%%%%%%%%%%%%%%%%%%%%%%%%%%%%%%%%%%%%%%%%%%%%%%
\ack
I am grateful to Erez Berg, Peter Hirschfeld, and Manfred Sigrist for helpful discussions. I owe special thanks to Eun-Ah Kim for fruitful discussions and help with the preparation of the manuscript. I acknowledge support from NSF Grant DMR-0955822 and NSF Grant DMR-1120296 to the Cornell Center for Materials Research. 

%%%%%%%%%%%%%%%%%%%%%%%%%%%%%%%%%%%%%%%%%%%%%%%%
% Appendix
%%%%%%%%%%%%%%%%%%%%%%%%%%%%%%%%%%%%%%%%%%%%%%%%
\appendix
\section{Transformation of the Basis States}
\label{app:transform2}
\begin{table}[tb!]
    \centering
\begin{tabular}{c||c c c c c|c c c c c}
  \hline\hline
  IR& E & 2S$_{4}$ & C$_{2}$ & 2C$_2'$ & 2$\sigma_{d}$ & $I$ & 2C$_4$ & $\sigma_h$ & 2$\sigma_v$ & 2C$_2''$\\
  \hline
  $A_{1g}$ \;\;& 1 & 1 & 1 & 1 & 1& 1 & 1 & 1 & 1 & 1\\[0.5ex] 
  $A_{2g}$ \;\;& 1 & 1 & 1 & -1 & -1& 1 & 1 & 1 & -1 & -1\\[0.5ex] 
  $B_{1g}$ \;\;& 1 & -1 & 1 & 1 & -1& 1 & -1 & 1 & 1 & -1\\[0.5ex] 
  $B_{2g}$ \;\;& 1 & -1 & 1 & -1 & 1& 1 & -1 & 1 & -1 & 1\\[0.5ex] 
  $E_g$ \;\;& 2 & 0 & -1 & 0 & 0& 2 & 0 & -1 & 0 & 0\\[0.5ex] 
  $A_{1u}$ \;\;& 1 & -1 & 1 & 1 & -1& -1 & 1 & -1 & -1 & 1\\[0.5ex] 
  $A_{2u}$ \;\;& 1 & -1 & 1 & -1 & 1& -1 & 1 & -1 & 1 & -1\\[0.5ex] 
  $B_{1u}$ \;\;& 1 & 1 & 1 & 1 & 1& -1 & -1 & -1 & -1 & -1\\[0.5ex] 
  $B_{2u}$ \;\;& 1 & 1 & 1 & -1 & -1& -1 & -1 & -1 & 1 & 1\\[0.5ex] 
  $E_u$ \;\;& 2 & 0 & -1 & 0 & 0& -2 & 0 & 1 & 0 & 0\\[0.5ex] 
  \hline\hline
\end{tabular}
\caption{Character table for $D_{4h}=D_{2d}\otimes I$. Note that the symmetry operations are ordered such that the first five columns belong to $D_{2d}$.}
    \label{tab:character}
\end{table}

In this appendix, we explicitly analyze how the symmetry transformations of the generating point group of the Fe-As layer, $D_{4h}$, act on the basis states 
\begin{equation}
  \vec{\mathcal{C}}_{\kk s} = (d^{(xz)}_{\kk s}, d^{(yz)}_{\kk s}, d^{(xz)}_{\kk+\Q s}, d^{(xy)}_{\kk+\Q s})^{T}
  \label{eq:vec}
\end{equation}
of the two-orbital model introduced in Section \ref{sec:2b}. 
For this purpose, we first analyze the behavior under all the symmetry operations of the unit cell $D_{2d}$ and than add inversion, since $D_{4h} = D_{2d}\otimes I$.
It is convenient to look at the respective states in real space, as depicted in Figure ~\ref{fig:unitcell}(b).
The two-fold rotation around the $z$ axis, $C_2$, amounts to multiplication of all the states by $-1$, i.e., the transformation matrix is given by $\hat{U}_{C_2}=-\hat{\Sigma}^{00}$. Therefore, all the basis matrices transform trivially under $C_2$, i.e. $\hat{U}_{C_2} \hat{\Sigma}^{ab} \hat{U}_{C_2}^{\dag}=\hat{\Sigma}^{ab}$. This already means that all possible orbital matrices only belong to one-dimensional representations of $D_{2d}$, as can be seen from the character table ~\ref{tab:character}.
We continue with the two-fold rotation around the $y$ axis. Under such a rotation, $d^{(xz)}_{is}$ ($\tilde{d}^{(xz)}_{is}$) stays the same, while $d^{(yz)}_{is}$ ($\tilde{d}^{(yz)}_{is}$) is multiplied by $-1$. The improper rotation $S_{4}$ is a combination of a $C_{4}$ rotation around $z$, followed by a mirror operation $\sigma_{h}$ at the $xy$ plane. This will take $d^{(xz)}_{is}$ to $d^{(yz)}_{is}$, $d^{(yz)}_{is}$ to $-d^{(xz)}_{is}$, as well as $\tilde{d}^{(xz)}_{is}$ to $-\tilde{d}^{(yz)}_{is}$ and $\tilde{d}^{(yz)}_{is}$ to $-\tilde{d}^{(xz)}_{is}$. Finally, the mirror operation can be obtained by combining $C_2'$ and $S_4$. This then yields the following eight matrix representations, 
\begin{equation}
  \hat{U}_E = \hat{\Sigma}^{00}, \quad \hat{U}_{C_{2}} = -\hat{\Sigma}^{00}, \quad \hat{U}_{C_{2}'} = \pm \hat{\Sigma}^{30},\quad \hat{U}_{S_{4}} =  \pm i\hat{\Sigma}^{20},\quad \hat{U}_{\sigma_{d}} =\pm \hat{\Sigma}^{10}.
  \label{eq:d2dmatrices}
\end{equation}
By calculating how the matrices $\hat{\Sigma}^{ab} = \hat{\rho}_{a}\otimes\hat{\tau}_{b}$ transform under these transformations, we can thus analyze what irreducible representation of $D_{2d}$ they belong to. 

Finally, the Fe-As lattice possesses centers of inversion lying between neighboring Fe sites, which transform the basis as $\hat{U}_I = \hat{\Sigma}^{03}$. 
The transformation properties of the individual basis matrices under $D_{2d}$ and inversion leads to the classification shown in Table~\ref{tab:basism}.
Note that $\hat{\Sigma}^{ab}$ with $b=1,2$ are odd under inversion. As can be seen from Table~\ref{tab:character}, this means that for example $\Sigma^{01}$, which belongs to $A_1$ in $D_{2d}$, belongs to $B_{1u}$ in $D_{4h}$. This can be understood from the transformation behavior of the $\hat{\tau}$ matrices~\cite{fischer:2011b}: While $\hat{\tau}^0$ and $\hat{\tau}^3$ transform trivially ($A_{1g}$), the matrices $\hat{\tau}^1$ and $\hat{\tau}^2$ transform as $B_{1u}$ in $D_{4h}$. As shown in Table~\ref{tab:character}, this is the irreducible representation that transforms in the same way as $A_{1g}$ in $D_{2d}$. The total matrix $\hat{\Sigma}^{ab}$ then transforms as the product of the irreducible representation of the orbital part with $B_{1u}$.

\section{Gell-Mann Matrices}
\label{app:gell-mann}
For the three-orbital model used in Section \ref{sec:3b}, we introduce the Gell-Mann matrices:
\begin{eqnarray}
  \lambda_1 &=& \left(\begin{array}{ccc} 0 & 1 & 0 \\ 1 & 0 & 0 \\ 0 & 0 & 0 \end{array}\right) \quad \lambda_2 = \left(\begin{array}{ccc} 0 & -i & 0 \\ i & 0 & 0 \\ 0 & 0 & 0 \end{array}\right),\quad \lambda_3 = \left(\begin{array}{ccc} 1 & 0 & 0 \\ 0 & -1 & 0 \\ 0 & 0 & 0 \end{array}\right),\nonumber\\
    \lambda_4 &=& \left(\begin{array}{ccc} 0 & 0 & 1 \\ 0 & 0 & 0 \\ 1 & 0 & 0 \end{array}\right), \quad \lambda_5 = \left(\begin{array}{ccc} 0 & 0 & -i \\ 0 & 0 & 0 \\ i & 0 & 0 \end{array}\right),\quad \lambda_6 = \left(\begin{array}{ccc} 0 & 0 & 0 \\ 0 & 0 & 1 \\ 0 & 1 & 0 \end{array}\right),\nonumber\\ 
    \lambda_7 &=& \left(\begin{array}{ccc} 0 & 0 & 0 \\ 0 & 0 & -i \\ 0 & i & 0 \end{array}\right)\!, \quad\! \lambda_8 = \frac{1}{\sqrt{3}}\!\left(\begin{array}{ccc} 1 & 0 & 0 \\ 0 & 1 & 0 \\ 0 & 0 & -2 \end{array}\right)\!,\quad \!\lambda_0 = \left(\begin{array}{ccc} 1 & 0 & 0 \\ 0 & 1 & 0 \\ 0 & 0 & 1 \end{array}\right)\!.
  \label{eq:lambdas}
\end{eqnarray}
Note that the matrices $(2\lambda_0+\sqrt{3}\lambda_8)/3$, $\lambda_1$, $\lambda_2$, and $\lambda_3$ correspond to the ($2\times2$) identity and Pauli matrices of the two-band model and 
\begin{equation}
  \frac13 (\lambda_0-\sqrt{3}\lambda_8) = \left(\begin{array}{ccc} 0 & 0 & 0 \\ 0 & 0 & 0 \\ 0 & 0 & 1 \end{array}\right).
\end{equation}

\end{document}